Title: Drag force in a hot non-relativistic, non-commutative Yang-Mills
plasma
Authors: Kamal L Panigrahi and Shibaji Roy
Comments: 18 pages, no figures, v2: added references, v3: typos fixed and as a result some eqs. have been corrected,
version to appear in JHEP, v4: correct version (a wrong version was uploaded in v3)

We apply the standard technique of Null Melvin Twist to the
non-extremal (D1, D3) bound state configuration of type IIB string
theory. Under a particular decoupling limit, such configuration
represents the gravity dual of the non-relativistic,
non-commutative Yang-Mills theory at a finite temperature. We then
use the AdS/CFT and the string probe approach to compute the drag
force on an external quark moving through such a hot
non-relativistic, non-commutative YM plasma. We discuss various
limiting cases to show the interplay between the non-relativistic
as well as the non-commutative effect of the general drag force
expression.

\documentclass[12pt]{article}

\usepackage{amsfonts}
\usepackage[centertags]{amsmath}
\usepackage{amssymb}
\usepackage{cite}
\usepackage{psfrag}
\usepackage{graphicx}
\usepackage{bbm,bm}
\usepackage{epsfig, multicol}

\allowdisplaybreaks
\begin{document}

\topmargin 0pt
\oddsidemargin 0mm
\def\be{\begin{equation}}
\def\ee{\end{equation}}
\def\bea{\begin{eqnarray}}
\def\eea{\end{eqnarray}}
\def\ba{\begin{array}}
\def\ea{\end{array}}
\def\ben{\begin{enumerate}}
\def\een{\end{enumerate}}
\def\nab{\bigtriangledown}
\def\tpi{\tilde\Phi}
\def\nnu{\nonumber}
\newcommand{\eqn}[1]{(\ref{#1})}

\newcommand{\vs}[1]{\vspace{#1 mm}}
\newcommand{\dsl}{\pa \kern-0.5em /}
\def\a{\alpha}
\def\b{\beta}
\def\g{\gamma}\def\G{\Gamma}
\def\d{\delta}\def\D{\Delta}
\def\ep{\epsilon}
\def\et{\eta}
\def\z{\zeta}
\def\t{\theta}\def\T{\Theta}
\def\l{\lambda}\def\L{\Lambda}
\def\m{\mu}
\def\f{\phi}\def\F{\Phi}
\def\n{\nu}
\def\p{\psi}\def\P{\Psi}
\def\r{\rho}
\def\s{\sigma}\def\S{\Sigma}
\def\ta{\tau}
\def\x{\chi}
\def\o{\omega}\def\O{\Omega}
\def\k{\kappa}
\def\pa {\partial}
\def\ov{\over}
\def\nn{\nonumber\\}
\def\ud{\underline}
\begin{flushright}
%
\end{flushright}
\begin{center}
{\large{\bf Drag force in a hot non-relativistic,
non-commutative\\ Yang-Mills plasma}}

\vs{10}

{Kamal L. Panigrahi$^a$\footnote{E-mail: panigrahi@phy.iitkgp.ernet.in}
and Shibaji Roy$^b$\footnote{E-mail: shibaji.roy@saha.ac.in}}

\vs{4}

{\it $^a$ Department of Physics and Meteorology\\
and\\
Centre for Theoretical Studies\\
Indian Institute of Technology Kharagpur, Kharagpur 721302, India}

\vs{4}

{\it $^b$ Saha Institute of Nuclear Physics\\
1/AF Bidhannagar, Calcutta 700064, India\\}

\end{center}

\vs{15}

\begin{abstract}
We apply the standard technique of Null Melvin Twist to the
non-extremal (D1, D3) bound state configuration of type IIB string
theory. Under a particular decoupling limit, such configuration
represents the gravity dual of the non-relativistic,
non-commutative Yang-Mills theory at a finite temperature. We then
use the AdS/CFT and the string probe approach to compute the drag
force on an external quark moving through such a hot
non-relativistic, non-commutative YM plasma. We discuss various
limiting cases to show the interplay between the non-relativistic
as well as the non-commutative effect of the general drag force
expression.

\end{abstract}
\newpage
\tableofcontents

\section{Introduction} The AdS/CFT correspondence 
(or its generalizations)
\cite{Maldacena:1997re,Witten:1998qj,Gubser:1998bc,Aharony:1999ti}
holographically relates weakly coupled string theory in a
particular background to a strongly coupled ('t Hooft coupling
$\lambda = g_{\rm YM}^2 N \gg 1$) relativistic conformal field
theory (or gauge theory). This is useful as it gives a
computational handle on the otherwise hard to access strongly
coupled gauge theory from supergravity. Indeed, such a strong-weak
duality has been, in recent years, proved to be instrumental for a
better understanding of various transport properties of certain
strongly coupled systems like quark gluon plasma (QGP) produced in
heavy ion collision (for reviews see
\cite{Herzog:2006gh,Son:2007vk,Shuryak:2008eq,Gubser:2009md,
Brasoveanu:2009ky}). Apart from nuclear physics strongly coupled
CFT's also appear in atomic and condensed matter physics and more
recently the holographic ideas have been applied to these systems
as well. For example, peculiar strong coupling behavior like
quantum Hall effect, Nernst effect, high temperature
superconductors, and quantum phase transitions in certain strongly
correlated electron systems can be understood at least
qualitatively by using the holographic dual descriptions involving
gravity
\cite{Herzog:2007ij,Hartnoll:2007ih,Hartnoll:2008vx,Hartnoll:2008kx,
Gubser:2008wv,Nakano:2008xc,Albash:2008eh,KeskiVakkuri:2008eb,Herzog:2009xv}.

In the examples mentioned above the CFT's were mainly of relativistic in
nature. However, for the application to most condensed matter systems, it
is useful to find holographic descriptions of CFT's which are
non-relativistic \cite{Hagen:1972pd,Mehen:1999nd,Nishida:2007pj}.
These systems sometimes can be produced in the laboratory
and indeed there exist such a strongly coupled non-relativistic system,
namely, the cold fermions at unitarity (for review see \cite{Giorgini})
which can be understood
using gravity/NRCFT correspondence, if the proper gravity dual for this system
can be found 
\cite{Balasubramanian:2008dm,Son:2008ye,Herzog:2008wg,Adams:2008wt}.
Motivated by the possible realizations of strongly coupled
CFT's in the laboratory there have been attempts to construct the gravity
duals in the form of non-relativistic branes in string theory
\cite{Mazzucato:2008tr}. The near
horizon geometry of these branes will have the isometry same as the
non-relativistic conformal (Schrodinger) symmetry of the boundary theory
and using the gravity/NRCFT correspondence one
can get a handle on the non-relativistic strongly coupled CFT from the
weakly coupled string theory or gravity.

In this paper we construct the non-relativistic non-extremal (D1,
D3) bound state solution of type IIB string theory. We will use
the standard procedure of Null Melvin Twist
\cite{Herzog:2008wg,Gimon:2003xk,Adams:2008wt, Mazzucato:2008tr} to construct
such a solution. A particular low energy limit, known as the
decoupling limit, of a stack of coincident (D1, D3) brane bound
state system gives rise to a non-commutative Yang-Mills (NCYM) theory on
the boundary
\cite{Seiberg:1999vs,Maldacena:1999mh,Hashimoto:1999ut}. It is
known that the D1-branes in the world-volume of D3-branes in the
decoupling limit produces a large magnetic or $B$-field
asymptotically and this is the source of a space-space
non-commutativity in the world-volume directions of D3-brane
\cite{Seiberg:1999vs}. The same decoupling limit for the stack of
coincident non-relativistic (D1, D3) bound state system will give
rise to a non-relativistic, NCYM theory on the
boundary. We will compute the drag force
\cite{Gubser:2006bz,Herzog:2006gh} experienced by an external
quark moving through this background of hot non-relativistic,
NCYM plasma. In this picture an external quark is
represented by the end point of a fundamental string attached to
the boundary carrying a fundamental charge under a gauge group and
is infinitely massive
\cite{Maldacena:1998im,Rey:1998bq,Rey:1998ik,Brandhuber:1998bs}.
The external quark loses its energy as the string attached to it
trails back and imparts a drag force on it. We will compute this
drag force when the quark moves along one of the non-commutative
directions for a sufficiently long time. We find that when the
boundary theory is both non-relativistic and non-commutative, it
is difficult to write the expression of the drag force in a closed
form. So, we will get the expression in various limiting cases to
show the interplay of the non-relativistic and non-commutative
effect. When the parameter characterizing the non-commutativity is
small, we find that there is no upper bound for the velocity. On
the other hand, when the non-commutativity parameter is large the
velocity of the quark can not be arbitrarily large in contrast to
what is expected of a non-relativistic theory. We will express the
drag force in terms of the parameters of the YM theory, namely,
the 't Hooft parameter, the temperature, the chemical potential
and the non-commutativity parameter in the various limiting cases.
Finally, we will formally integrate the drag force expression to
compute the momentum or energy loss \cite{Gubser:2006bz} of the
quark moving in the hot non-relativistic NCYM theory.

This paper is organized as follows. In the next section we discuss the
Null Melvin Twist on the non-extremal (D1, D3) bound state system of 
type IIB string theory
and also the decoupling limit. In section 3, we calculate the drag force
on a heavy quark moving through the hot non-relativistic NCYM plasma and
discuss the various limits to understand the general drag force expression
in the various corners of the solution space. We conclude in section 4.

\section{Null Melvin Twist on non-extremal (D1, D3) 
bound state solution}

The non-extremal (D1, D3) bound state configuration of type IIB string theory
is given as \cite{Breckenridge:1996tt,Costa:1996zd,Harmark:1999rb},
\bea\label{d1d3}
ds^2 &=& H^{-\frac{1}{2}}\left[-f(dx^0)^2 + (dx^1)^2 + \left(\frac{H}{F}\right)
\left((dx^2)^2 + (dx^3)^2\right)\right]+ H^{\frac{1}{2}}\left[f^{-1}dr^2 + r^2
d\Omega_5^2\right]\nn
e^{2\phi} &=& g_s^2 \frac{H}{F}\nn
B_{[2]} &=& \tan \theta F^{-1} dx^2 \wedge dx^3, \qquad A_{[2]} =
-\frac{1}{g_s} \sin\theta \coth\varphi F^{-1} dx^0 \wedge dx^1\nn
F_{[5]} &=& -\frac{1}{g_s} \cos\theta \coth\varphi \left(\frac{H}{F}\right)
\partial_r H^{-1} dx^0\wedge
dx^1\wedge dx^2 \wedge dx^3 \wedge dr
\eea
where the various functions appearing in the solution \eqn{d1d3} are defined
as,
\be
f = 1 - \frac{r_0^4}{r^4}, \quad H = 1 +
\frac{r_0^4\sinh^2\varphi}{r^4},\quad F = 1 + \frac{r_0^4 \cos^2\theta
  \sinh^2\varphi}{r^4}
\ee
Note that the metric in the above is given in the string frame. The
dilaton $\phi$ is non-constant and $g_s$ is the string coupling. The 5-form
field strength\footnote{Note that the 5-form field
strength must be made self-dual by adding the hodge-dual $\ast F_{[5]}$
with the $F_{[5]}$ given above for \eqn{d1d3} to be a solution of type IIB
supergravity, although we have not written it explicitly.}
and $A_{[2]}$ tell us the presence of D3 and D1 branes in the
solution respectively. The non-zero NSNS $B_{[2]}$ field is due the
non-threshold nature of the bound state (D1, D3). It is clear from the
solution \eqn{d1d3} that D3-branes are lying along $x^1,\ldots,x^3$, whereas
D1-branes are lying along $x^1$. The angle $\theta$ measures the relative
numbers of D3-branes and D1-branes and is defined as, $\cos\theta =
N/\sqrt{N^2 + M^2}$, where $N$ is the number of D3-branes and $M$ is the
number of D1-branes per unit co-dimension two-volume transverse to the
D1-branes. Also in the above $\varphi$ is the boost parameter and $r_0$ is the
radius of the horizon of non-extremal or black (D1, D3)-brane solution.

Eq.\eqn{d1d3} represents the relativistic (D1, D3) solution. The
corresponding non-relativistic solution can be obtained by applying the
standard procedure of the so-called Null Melvin Twist 
\cite{Gimon:2003xk,Herzog:2008wg,Adams:2008wt,Mazzucato:2008tr}
or by taking a Penrose
limit or a TsT transformation \cite{Maldacena:2008wh} to the relativistic 
(D1, D3) solution. The
procedure of Null Melvin Twist generates a new solution in eight steps
starting from the original relativistic solution.
So, starting from the black
(D1, D3) bound state solution given in \eqn{d1d3}, we first apply boost along
$x^1$-direction, T-dualize the boosted isometric $x^1$-direction, twist a
local one-form in a transverse compact direction, T-dualize back along $x^1$,
boost back along $x^1$ and then take a scaling limit. In the light-cone
coordinates $t = (x^1 + x^0)/\sqrt{2},\,\, \xi = (x^1 - x^0)/\sqrt{2}$
the final solution takes the form,
\bea\label{nrd1d3}
ds^2 &=& \frac{H^{-\frac{1}{2}}}{K}\left[\left\{-\left(2 r^2 \beta^2 f +
      \frac{g}{2} \right)dt^2 - \frac{g}{2} d\xi^2 + \left(1 + f\right)dt
d\xi\right\} \right.\nn
& & \left. +\left(\frac{H}{F}\right) K \left((dx^2)^2+(dx^3)^2
\right)\right] + H^{\frac{1}{2}}\left[f^{-1} dr^2 + r^2 \left(\frac{1}{K}
\left(d\chi + {\cal A}\right)^2 + ds_{\rm P^2}^2\right)\right]
\eea
Here $K = 1 - \beta^2 r^2 g(r)$ and $g(r) = - r_0^4/r^4$. We have introduced
a one form ${\cal A}$ by $d{\cal A} = J$, with $J$, the Kahler form on the
complex projective space ${\rm P^2}$ and $ds_{\rm P^2}^2$ is the metric on the
complex projective space ${\rm P^2}$. The part of the metric
$(1/K)(d\chi + {\cal A})^2 + ds_{\rm P^2}^2$ is the metric on the squashed
5-sphere with the squashing parameter $K$. The other fields are given as,
\bea\label{nrother}
e^{2\phi} &=& g_s^2 \left(\frac{H}{F}\right)\frac{1}{K}\nn
B_{[2]} &=& \frac{r^2\beta}{\sqrt{2} K} \left(d\chi + {\cal A}\right) \wedge
\left[(1+f) dt + (1-f) d\xi\right] + \frac{\tan\theta}{F} dx^2 \wedge dx^3\nn
A_{[2]} &=& \frac{1}{g_s F} \sin\theta \coth\varphi dt \wedge d\xi\nn
F_{[5]} &=& F_{[5]}^{({\rm old})} + \frac{1}{2}\left(B_{[2]}^{({\rm old})}
\wedge dA_{[2]}^{({\rm old})} - A_{[2]}^{({\rm old})}
\wedge dB_{[2]}^{({\rm old})}\right)\nn
& & \quad - \frac{1}{2}\left(B_{[2]}^{({\rm new})}
\wedge dA_{[2]}^{({\rm new})} - A_{[2]}^{({\rm new})}
\wedge dB_{[2]}^{({\rm new})}\right)
\eea
where $F_{[5]}^{({\rm old})},\, B_{[2]}^{({\rm old})},\, A_{[2]}^{({\rm old})}$
are the various fields given in \eqn{d1d3} and $B_{[2]}^{({\rm new})},\,
A_{[2]}^{({\rm new})}$ are the fields given in \eqn{nrother}. The solution
represented by \eqn{nrd1d3} and \eqn{nrother} are the non-relativistic
non-extremal (D1, D3) solution of type IIB string theory. The extremal
solution with the non-relativistic symmetry can be obtained from
\eqn{nrd1d3} and \eqn{nrother} by scaling $r_0 \to 0$ and
$\varphi \to \infty$ such
that the product $r_0^2 \sinh\varphi$ remains finite. Note that the parameter
$\beta$, appearing in the solution above can be scaled away in the extremal
case by scaling the $t$ and $\xi$
coordinates appropriately. However, this can not be done for the non-extremal
case and in this case $\beta$ is a physical parameter related to the chemical
potential of the boundary theory \cite{Adams:2008wt}.

The non-commutative Yang-Mills (NCYM) decoupling limit
\cite{Maldacena:1999mh,Hashimoto:1999ut} is a low energy limit
by which we zoom into the region,
\be\label{zoom}
r_0 < r \sim r_0 \sinh^{\frac{1}{2}}\varphi \cos^{\frac{1}{2}}\theta \ll
r_0 \sinh^{\frac{1}{2}}\varphi
\ee
Note that in this region $\varphi$ is very large, whereas, the angle $\theta$
is very close to $\pi/2$. In this approximation,
\be\label{approx}
H \approx \frac{r_0^4 \sinh^2\varphi}{r^4}, \qquad \frac{H}{F} \approx
\frac{1}{\cos^2\theta(1 + a^4 r^4)} \equiv \frac{h}{\cos^2\theta}
\ee
where,
\be\label{def}
h = \frac{1}{1+a^4r^4}, \quad {\rm with} \quad a^4 =
\frac{1}{r_0^4\sinh^2\varphi \cos^2\theta}
\ee
Note that in the NCYM limit, the asymptotic value of $B_{23}$ component
(from \eqn{nrother} we find that this is $\tan\theta$) responsible for
creating non-commutativity takes a very large value as $\theta \to \pi/2$.
Now with the above approximation \eqn{approx}, the metric in \eqn{nrd1d3}
takes the form,
\bea\label{nrd1d3a}
ds^2 &=& \frac{r^2}{R^2} \frac{1}{K}\left[\left\{-\left(2r^2\beta^2 f +
        \frac{g}{2} \right)dt^2 - \frac{g}{2} d\xi^2 + (1 +
      f)dtd\xi\right\}\right.\nn
& & \left. + h K\left((dx^2)^2 + (dx^3)^2\right)\right] + \frac{R^2}{r^2}\left[
f^{-1} dr^2 + r^2 \left(\frac{1}{K}\left(d\chi + {\cal A}\right)^2 +
ds_{\rm P^2}^2\right)\right]
\eea
where $R^2 = r_0^2\sinh\varphi$ and we have rescaled the coordinates
$x^{2,3}$ as $x^{2,3} \to \cos\theta x^{2,3}$. Due to the large magnetic
or $B$-field in $x^{2,3}$-directions, they satisfy the non-commutativity
relation $[x^2, x^3] = i\Theta$, where $\Theta$ is the non-commutativity
parameter.
Similarly the other fields\footnote{Since we do not need the other fields in
what follows we will not write them explicitly here.} in \eqn{nrother}
can also be rewritten using
\eqn{approx}, and this will be the holographic dual of
non-relativistic NCYM theory. Note that since for the non-relativistic case
$\beta$ is a physical parameter, by setting $\beta$ to zero, we recover the
near horizon metric of the relativistic non-extremal (D1, D3)
solution. However, for the extremal case, $\beta$ can not be put to zero, but
should be scaled away before recovering the relativistic limit.

\section{Drag force in hot non-relativistic NCYM plasma}

Now in order to compute the drag force\footnote{Drag force in various 
other backgrounds
have been calculated earlier in 
\cite{Fadafan:2008gb,Fadafan:2008uv,Sadeghi:2008ws,Sadeghi:2009mp,
Sadeghi:2009hh,Fadafan:2009an}.} on an external quark it is convenient
to write the metric \eqn{nrd1d3a} in the original coordinate
\cite{Akhavan:2008ep}
$t \to (\xi - t)/\sqrt{2}$ and $\xi \to (\xi + t)/\sqrt{2}$ as,
\bea\label{nrd1d3b}
ds^2 &=& \frac{r^2}{R^2} \frac{1}{K}\left[\left\{-\left(1 +
        r^2 \beta^2 \right) f dt^2 + \left(1 - r^2 \beta^2 f\right) d\xi^2
+ 2 r^2 \beta^2 f dtd\xi\right\}\right.\nn
& & \left. + h K\left((dx^2)^2 + (dx^3)^2\right)\right] + \frac{R^2}{r^2}\left[
f^{-1} dr^2 + r^2 \left(\frac{1}{K}\left(d\chi + {\cal A}\right)^2 +
ds_{\rm P^2}^2\right)\right]
\eea
It is clear from \eqn{nrd1d3b} that by putting $\beta$ to zero we recover the
relativistic limit as expected. Now the dynamics of an external quark moving
in this background can be understood from the Nambu-Goto action of the
fuandamental string given as,
\be\label{ngaction}
S = -\frac{1}{2\pi\alpha'}\int d\tau d\sigma \sqrt{-{\rm det}(g_{ab})}
\ee
where $g_{ab}$ is the induced metric on the world-sheet of the fundamental
string in the background \eqn{nrd1d3b} and is given as,
\be
g_{ab} = \frac{\partial X^\mu}{\partial \xi^a} \frac{\partial X^\nu}
{\partial \xi^b} G_{\mu\nu}
\ee
where $G_{\mu\nu}$ is the background metric \eqn{nrd1d3b} and $\xi^{a,b}$,
$a,b=0,1$ are the world-sheet coordinates $\tau = \xi^0$ and $\sigma = \xi^1$.
We now use the static gauge condition $X^0 \equiv t = \tau$ and $r=\sigma$.
The end point of the string is allowed to move along one of the
non-commutative directions $X^2 = x$ (say). Then the string embedding is
completely specified by the function $x(t,r)$. The action \eqn{ngaction}
then reduces to the form,
\be\label{ngreduced}
S = -\frac{1}{2\pi\alpha'}\int dt dr \left[\frac{1+r^2\beta^2}{K} +
\left(\frac{r^2}{R^2}\right)^2 \frac{1+r^2\beta^2}{K} h f (x')^2 - \frac{h}{f}
(\dot x)^2\right]^{\frac{1}{2}}
\ee
Here `overdot' and `prime' on $x$ denote the derivative with respect to `$t$'
and `$r$' respectively. Let us now make a simplifying and reasonable
assumption that if we allow a sufficiently long time the string will move
with a constant velocity and therefore, $x(t,r) = v t +
\zeta(r)$. Substituting this in \eqn{ngreduced}, the string action takes the
form,
\be\label{ngreduceda}
S = -\frac{1}{2\pi\alpha'}\int dt dr \left[\frac{1+r^2\beta^2}{K} +
\left(\frac{r^2}{R^2}\right)^2 \frac{1+r^2\beta^2}{K} h f (\zeta')^2 -
\frac{h}{f} v^2\right]^{\frac{1}{2}}
\ee
Since the Lagrangian density does not contain $\zeta$ explicitly,
the corresponding
momentum must be conserved independent of both $r$ and $t$ and so, we have
\be\label{conserved}
\pi_{\zeta} = \frac{\left(\frac{r^4}{R^4}\right) \frac{1+r^2\beta^2}{K}
h f \zeta'}{\sqrt{\frac{1+r^2\beta^2}{K} + \left(\frac{r^4}{R^4}\right)
\frac{1+r^2\beta^2}{K} h f \left(\zeta'\right)^2 - \frac{h}{f} v^2}} =
{\rm const.\,\, independent\,\,of\,\,} r, t
\ee
Solving this equation we obtain,
\be\label{solution}
\zeta' = \frac{R^4}{r^4}\frac{\pi_{\zeta} \sqrt{K}}{\sqrt{1 + r^2 \beta^2}
h f} \sqrt{\frac{\frac{f}{h}\frac{1+r^2\beta^2}{K} - v^2}
{\frac{f}{h}\frac{1+r^2\beta^2}{K} - \frac{\pi_{\zeta}^2 R^4}{r^4 h^2}}}
\ee
Now notice in \eqn{solution} that as $r$ varies from $r_0$ to $\infty$, both
the numerator and the denominator inside the square root change sign. So, at
some $r$, the expression for $\zeta'$ can become imaginary when they have
opposite signs. Therefore, 
the solution
\eqn{solution} is not always physically acceptable. To get the physical
solution we have to choose the constant $\pi_{\zeta}$ suitably
such that both the numerator and the denominator in the square root change
the sign at the same place $r_v$ (say). This fixes the constant $\pi_{\zeta}$
in the form,
\be\label{momentum}
\pi_{\zeta} = \frac{r_v^2}{R^2}\frac{v}{(1+a^4r_v^4)} = \frac{r_v^2}{r_0^2}
\frac{v}{\sinh\varphi (1+a^4r_v^4)}
\ee
where $r_v$ is the solution of the equation given by,
\be\label{rveqn}
(1+ r_v^2\beta^2)\left[a^4 r_v^8 + (1-a^4r_0^4 - v^2) r_v^4 - r_0^4\right]
+ v^2 \beta^2 r_v^2\left(r_v^4 - r_0^4\right) = 0
\ee
We can substitute the value of $\pi_{\zeta}$ from \eqn{momentum} to
\eqn{solution} and integrate to obtain the complete string dynamics. In
principle this is possible, but in practice, the difficulty is that
the eq.\eqn{rveqn} is a polynomial equation in $r_v$ of degree ten and in
general it is not always possible to solve it analytically. Even if this
is possible it is not always guaranteed that the eq.\eqn{solution}
can be integrated in a closed form. However, we can formally write down
the expression of the drag force from \eqn{momentum} as,
\be\label{dragforce}
F = -\frac{1}{2\pi\alpha'} \frac{r_v^2}{R^2}\frac{v}{(1+a^4r_v^4)} =
-\frac{1}{2\pi\alpha'}\frac{r_v^2}{r_0^2} \frac{v}{\sinh\varphi (1+a^4r_v^4)}
\ee
Note that by solving $r_v$ from eq.\eqn{rveqn}, we can express the drag force
\eqn{dragforce} in terms of $r_0$, $\sinh\varphi$, $v$, $\beta$, $a$
and $\alpha'$, the
parameters of the string theory or gravity side. Later we will express the
drag force expression in terms of the parameters of the boundary gauge theory
or non-relativistic NCYM theory. Let us mention here that the parameter
$\beta$ is non-trivial and cannot be scaled away for non-relativistic theory
and therefore we will call it the non-relativistic parameter and by setting
it to zero, we can recover the relativistic limit. On the other hand the
parameter $a$, is associated with the non-commutativity of the theory and by
setting it to zero we can recover the commutative limit. The parameter $a$ was
defined before as $a^4 r_0^4 = 1/(\sinh^2\varphi \cos^2\theta)$. In the NCYM
decoupling limit $\sinh\varphi \sim 1/\alpha'$ and
$\cos\theta \sim \alpha'/\Theta$, where $\Theta$ is the non-commutativity
parameter given earlier and so, as $\alpha' \to 0$,
$\sinh\varphi \gg 1$ and $\cos\theta \to 0$ as we mentioned
earlier. Therefore, in this decoupling limit $a^4 r_0^4 \sim \Theta^2$. So,
$a$ measures the non-commutativity as it is directly related to the
non-commutativity parameter.

In order to understand the drag force expression \eqn{dragforce} more
concretely we make some observation from the $r_v$ equation given in
\eqn{rveqn}. We note that for the relativistic ($\beta = 0$) and
commutative ($a=0$)  case, we get from \eqn{rveqn}
\be\label{relcom}
v^2 = \frac{r_v^4 - r_0^4}{r_v^4}
\ee
It is therefore clear that as $r_v$ goes from $r_0$ to $\infty$, $v$ varies
from 0 to 1 as expected of a relativistic theory. On the other hand, for the
commutative ($a=0$) but non-relativistic ($\beta \neq 0$) case, we get from
\eqn{rveqn},
\be\label{nonrelcom}
v^2 = \frac{(1+r_v^2\beta^2)(r_v^4 - r_0^4)}{r_v^2(r_v^2 + r_0^4 \beta^2)}
\ee
In this case, as $r_v$ varies from $r_0$ to $\infty$, $v$ varies from
0 to $\infty$, again this is as expected of a non-relativistic theory.
Let us next consider the relativistic ($\beta = 0$) but non-commutative
($a \neq 0$) case. We find from \eqn{rveqn},
\be\label{relnoncom}
v^2 = \frac{(r_v^4 - r_0^4)(a^4 r_v^4 + 1)}{r_v^4}
\ee
Here we notice that for $r_v=r_0$, $v=0$. But since the maximum value $v$ can
take for a relativistic theory is 1, $r_v$ can not be arbitrarily large. We
can calculate the value of $r_v$, when $v=1$ from eq.\eqn{relnoncom} and we
find (for large and small non-commutativity)
\bea\label{rvvalue1}
a^4 r_v^4 & = & a^4 r_0^4 + 1 - \frac{1}{a^4 r_0^4} +  \cdots, \qquad\qquad
{\rm when} \quad ar_0 \gg 1\\\label{rvvalue2}
a^4 r_v^4 & = & a^2 r_0^2 + \frac{1}{2} a^4 r_0^4 + \cdots, \qquad
{\rm when} \quad ar_0 \ll 1
\eea
So, for large non-commutativity $r_v$ is close to $r_0$, but for small
non-commutativity $r_v$ is far away from $r_0$. Finally, we consider both
non-relativistic ($\beta \neq 0$) and non-commutative ($a \neq 0$) case.
We find from \eqn{rveqn},
\be\label{nonrelnoncom}
v^2 = \frac{(1+ r_v^2 \beta^2)(r_v^4 - r_0^4)(a^4 r_v^4 + 1)}
{r_v^2(r_v^2 + r_0^4 \beta^2)}
\ee
Here also we note that as $r_v$ starts from $r_0$, $v$ starts from 0. We just
mentioned that $r_v$ can not take arbitrary large value when $a\neq 0$, for
$v$ to remain less than or equal to 1 (for the relativistic theory). However,
we will see that even when $\beta \neq 0$ (i.e. for the non-relativistic
theory) $r_v$ can not take arbitrary large value. The reason is that if $r_v$
exceeds the value obtained for the relativistic case given in
eqs.\eqn{rvvalue1}, \eqn{rvvalue2}, then $v$ will exceed one when we put
$\beta = 0$ and this will be unphysical for a relativistic theory. Therefore, 
we will use the values of $r_v$ given in \eqn{rvvalue1} and \eqn{rvvalue2} to
determine $v$ when $\beta \neq 0$. It can be checked from \eqn{nonrelnoncom}
that for $ar_0\ll 1$, $v$ can be much larger, i.e., $v \gg 1$ (showing the
non-relativistic nature of the theory), but for $ar_0 \gg 1$, the maximum
value of $v$ is of the order 1. Indeed it can be checked from \eqn{rvvalue1}
and \eqn{nonrelnoncom} that the value of $v$ is given by,
\be\label{noncomv}
v^2 = 1 + \frac{r_0^2\beta^2}{a^4 r_0^4 \left(1+r_0^2\beta^2\right)} +
O(\frac{1}{a^8 r_0^8})
\ee
So, for $a^4 r_0^4 \gg 1$, the velocity of the quark $v$ is close to
1 (but the velocity is always greater than 1) as we see from \eqn{noncomv}. 
So, we will analyse the general $r_v$ equation
\eqn{rveqn} and the drag force \eqn{dragforce} in four different cases,
namely, $(i)$ $v \ll 1,\,\,ar_0 \ll 1$, $(ii)$ $v\ll 1,\,\,ar_0 \gg 1$,
$(iii)$ $v\gg 1,\,\,ar_0 \ll 1$, and $(iv)$ $v \sim 1,\,\,ar_0 \gg 1$.

$(i)$ $v \ll 1,\,\,ar_0 \ll 1$. In this case \eqn{rveqn} can be solved to
obtain $r_v$ in the following form,
\be\label{casearv}
a^4 r_v^4 = a^4 r_0^4 (1 + v^2 + v^4) - a^8 r_0^8 v^2 (1+v^2) + \cdots
\ee
Substituting this in \eqn{dragforce} we obtain
\be\label{caseadf}
F = - \frac{1}{2\pi\alpha'}\frac{v}{R^2}r_0^2
\left[\left(1+\frac{1}{2}v^2\right) - \left(1 + 2 v^2\right) a^4 r_0^4 +
    \cdots\right]
\ee
This matches exactly with those given in refs.\cite{Matsuo:2006ws,Roy:2009sw}
with $v \ll 1$ as it
should be in this approximation.

$(ii)$ $v \ll 1,\,\,ar_0 \gg 1$. In this case by solving \eqn{rveqn} we obtain
$r_v$ in the form,
\be\label{casebrv}
a^4 r_v^4 = a^4 r_0^4\left[1 + \frac{v^2}{a^4r_0^4} + \cdots\right]
\ee
Substituting this is \eqn{dragforce} we obtain,
\be\label{casebdf}
F = -\frac{1}{2\pi\alpha'} \frac{v}{R^2}\frac{r_0^2}{a^4 r_0^4}\left[1 -
\frac{v^2 + 2}{2a^4r_0^4} + \cdots\right]
\ee
This also matches with the drag force expression given in
\cite{Matsuo:2006ws,Roy:2009sw} with
$v \ll 1$ as expected.

$(iii)$ $v \gg 1,\,\,ar_0 \ll 1$. In this case we get from \eqn{rveqn}
\be\label{casecrv}
a^4 r_v^4 = a^2 r_0^2\left(1 + \frac{a^2 r_0^2}{2} + \cdots\right)
\ee
Substituting these in \eqn{dragforce} we get,
\be\label{casecdf}
F = -\frac{1}{2\pi\alpha'}\frac{v}{R^2 \beta^2}\left(v^2 + r_0^4
  \beta^4\right) \left(1-\frac{3}{4} a^2 r_0^2 + \cdots\right)
\ee

$(iv)$ $v \sim 1,\,\,ar_0 \gg 1$. Eq.\eqn{rveqn} in this case gives,
\be\label{casedrv}
a^4 r_v^4 = a^4 r_0^4 \left(1 + \frac{1}{a^4r_0^4} - \frac{1}{a^8 r_0^8} 
\cdots\right)
\ee
Substituting this in \eqn{dragforce} we get,
\be\label{caseddf}
F = -\frac{1}{2\pi\alpha'} \frac{v(v^2-1)}{R^2} 
\frac{(1+r_0^2\beta^2)}{\beta^2} \left(1 -
  \frac{3}{2}\frac{1}{a^4r_0^4} + \cdots\right)
\ee
In eqs.\eqn{caseadf}, \eqn{casebdf}, \eqn{casecdf} and \eqn{caseddf} we
have given the drag force expressions in terms of the parameters of string
or gravity theory. Now in order to understand the nature of the force in terms
the boundary gauge theory, we have to relate the gravity parameters with the
parameters of the non-relativistic NCYM theory. The temperature of the
non-relativistic NCYM theory can be calculated from the Hawking temperature of
the non-extremal decoupled gravity configuration given in \eqn{nrd1d3b} and
has the form,
\be\label{temp}
T = \frac{1}{\pi r_0 \sinh\varphi}
\ee
Also from the charge of the D3-brane we can calculate,
\be\label{relation}
r_0^4 \sinh^2\varphi = 2 \hat{\lambda} \alpha'^2,
\ee
where $\hat{\lambda} = \hat{g}_{YM}^2 N$, is the 't Hooft coupling of the
non-relativistic NCYM theory, $\hat{g}_{YM}$ is the NCYM coupling and $N$
is the number of D3-branes and is related to the gauge group $SU(N)$ of the
gauge theory. The 't Hooft parameter $\hat\lambda$ is related to the
corresponding parameter of ordinary YM theory by the scaling of the form,
$\lambda = (\alpha'/\Theta)\hat\lambda$, where $\Theta$ is the
non-commutativity parameter \cite{Maldacena:1999mh,Hashimoto:1999ut}.
Now using \eqn{temp} and \eqn{relation} we get,
\be\label{relation1}
\sinh\varphi = \frac{1}{\sqrt{2\hat\lambda} \pi^2 T^2 \alpha'}, \quad
r_0 = \sqrt{2\hat\lambda} \pi T \alpha', \quad {\rm and}, \quad
a^4 r_0^4 = 2\hat\lambda \pi^4 T^4 \Theta^2
\ee
In obtaining the last expression we have used $a^4r_0^4 =
1/(\sinh^2\varphi\cos^2\theta)$ and $\cos\theta = \alpha'/\Theta$.
Using \eqn{relation1} we will express the drag force given earlier in
\eqn{caseadf}, \eqn{casebdf}, \eqn{casecdf}, \eqn{caseddf} for various cases
in the leading order in terms of the non-relativistic NCYM theory.

In the first case $(i)$ $v \ll 1,\,\,ar_0 \ll 1$, we get
\be\label{caseadfYM}
F = - \sqrt{\frac{\hat{g}_{YM}^2 N}{2}} \pi T^2 v
\ee
We can formally express the above expression \eqn{caseadfYM} in terms of
the momentum $p$ and mass $m$ of the external quark and integrate to find
\cite{Gubser:2006bz,Akhavan:2008ep},
\be\label{caseamom}
p(t) = p(0) e^{-\pi\sqrt{\frac{\hat{g}_{YM}^2 N}{2}}\frac{T^2}{m}t},
\qquad {\rm where} \quad
p(0) \ll m
\ee
The corresponding energy will be given as,
\be\label{caseaener}
E(t) = E(0) e^{-\pi\sqrt{2\hat{g}_{YM}^2 N}\frac{T^2}{m}t},
\qquad {\rm where} \quad
E(0) \ll m/2
\ee

The expression of drag force for the case  $(ii)$ $v \ll 1,\,\,ar_0 \gg 1$
can be written using \eqn{relation1} in the leading order as,
\be\label{casebdfYM}
F = -\frac{1}{2\sqrt{2 \hat{g}_{YM}^2 N} \pi^3 T^2} \frac{v}{\Theta^2}
\ee
The momentum and energy can be obtained as before and have the forms,
\bea\label{casebmom}
p(t) &=& p(0) e^{-\frac{t}{2\pi^3 m T^2 \sqrt{2\hat{g}_{YM}^2 N} \Theta^2}},
\qquad {\rm where}\quad p(0)
\ll m \\\label{casebener}
E(t) &=& E(0) e^{-\frac{t}{\pi^3 m T^2 \sqrt{2\hat{g}_{YM}^2 N}\Theta^2}},
\qquad {\rm where} \quad
E(0) \ll m/2
\eea

In the above two cases when $v \ll 1$, that is, when the momentum or energy
is much less than the quark mass, the quark will lose its momentum or energy
exponentially. The relaxation times are different in the two different cases
and depend on whether the non-commutativity is small or large. For small
non-commutativity the relaxation time does not depend on the non-commutativity
parameter in the leading order, but depend directly on the mass of the quark
and inversely on the square of the temperature as well as the square-root of
the 't Hooft coupling. On the other hand, for large non-commutativity, the
dependence on the temperature and the 't Hooft coupling get inverted, but the
dependence on the mass remains the same. Also, in this case, the relaxation
time depends directly on the square of the non-commutativity parameter. So,
for small non-commutativity the non-commutative effect does not show up in the
leading order, but it does show up in the leading order for large
non-commutativity.

Similarly, for case $(iii)$ $v \gg 1,\,\,ar_0 \ll 1$ using \eqn{relation1}
the expression of drag force in the leading order has the form
\be\label{casecdfYM}
F = - \frac{v^3 \mu^2}{2\pi \sqrt{2 \hat{g}_{YM}^2 N}}
\ee
where we have defined $1/(\beta \alpha') = \mu$, the chemical potential of the
non-relativistic NCYM theory. The momentum and the energy in this case have the
forms,
\bea\label{casecmom}
p(t) &=& \left[\frac{1}{p(0)^2} + \frac{t \mu^2}{\pi m^3\sqrt{2
\hat{g}_{YM}^2 N}}\right]^{-\frac{1}{2}},\qquad
{\rm where} \quad p(0) \gg m\\
\label{casecener}
E(t) &=& \left[\frac{1}{E(0)} + \frac{2t\mu^2}{\pi m^2
\sqrt{2\hat{g}_{YM}^2 N}}\right]^{-1},\qquad
{\rm where} \quad E(0) \gg m/2
\eea
In this case the momentum or the energy loss does not
depend on the temperature \cite{Akhavan:2008ep} unlike
in the previous two cases. The momentum (or the energy) loss depends directly
on the square of the chemical potential and inversely on the cube (square)
of the quark mass and the square-root of the 't Hooft coupling. With time
they do not decay exponentially as in the previous cases, but the momentum
goes as $t^{-1/2}$, whereas the energy goes as $t^{-1}$.

Finally, the drag force expression for case $(iv)$ $v \sim 1,\,\,ar_0 \gg 1$
can be written using \eqn{relation1} in the leading order as,
\be\label{caseddfYM}
F = -\sqrt{\frac{\hat{g}_{YM}^2 N}{2}} \pi T^2 C(\hat{g}_{YM}^2 N, \mu,T)(v^3-v)
\ee
where 
\be\label{cfunction}
C(\hat{g}_{YM}^2 N, \mu, T) = \frac{\mu^2 + 2\pi^2 \hat{g}_{YM}^2 N T^2}
{2 \pi^2 \hat{g}_{YM}^2 N T^2}
\ee
with $\mu$ being the chemical potential defined before. Note that the 
function $C$ tends to unity when $\mu^2 \ll 2 \pi^2\hat{g}_{YM}^2 N T^2$. Also
note that when $\mu^2 \gg 2 \pi^2 \hat{g}_{YM}^2 N T^2$, the force expression
is independent of temperature. 
The momentum and energy in this case have the forms,
\bea\label{casedmom}
p(t) &=& m \left[1 - \left(1-\frac{m^2}{p^2(0)}\right)e^{-\frac{2\pi T^2}{m}
\sqrt{\frac{\hat{g}_{YM}^2 N}{2}} C(\hat{g}_{YM}^2 N, \mu, T) t}
\right]^{-\frac{1}{2}},
\,\,\, {\rm where} \,\,\, p(0) \sim m\\
\label{casedener}
E(t) &=& \frac{m}{2} \left[1 - \left(1-\frac{m}{2E(0)}\right)
e^{-\frac{2\pi T^2}{m}
\sqrt{\frac{\hat{g}_{YM}^2 N}{2}} C(\hat{g}_{YM}^2 N, \mu, T) t}
\right]^{-1},
\,\,\,{\rm where}\,\,\, E(0) \sim \frac{m}{2}
\eea
In this case since $v \sim 1$, both the momentum and the energy loss is very
small due to large non-commutative effects.

\section{Conclusion}

To summarize, in this paper we have considered the non-extremal
(D1, D3) bound state solution of type IIB string theory. In a
particular decoupling limit this supergravity configuration is
known to describe the holographic dual of relativistic NCYM theory
at finite temperature with space-space non-commutativity. By
applying the standard technique of Null Melvin Twist we have
obtained the non-relativistic version of non-extremal (D1, D3)
bound state system. The same decoupling limit in this case,
describes the holographic dual of the non-relativistic NCYM theory
at finite temperature with space-space non-commutativity. We have
computed the drag force on a quark moving through such plasma and
along one of the non-commutative directions, by using the AdS/CFT
correspondence and the string probe approach. We first computed
the drag force in terms of the parameters of the gravity theory
and then using the AdS/CFT dictionary expressed it in terms of the
parameters of the gauge theory. We found that the general drag
force expression can not be written in a closed form. So, to show
the various effects we have considered the various corners of the
solution space and obtained the drag force expressions in the
leading order. We also formally integrated the drag force
expression to obtain the momentum as well as the energy loss of
the quark in various limits. In particular, we have shown that
when the velocity of the quark is small, it loses its energy
exponentially with time. The relaxation times are expressed in
terms of the parameters of the NCYM theory. When the
non-commutative effect is large, the relaxation time depends
explicitly on the non-commutativity parameter. We also found that
the velocity could be very large when the non-commutative effect
is small. In that case the quark loses its energy as inverse power
of time. Also the energy loss does not depend on the temperature
of the theory unlike in other cases. Finally, we found that when
the non-commutative effect is large the velocity can not be
arbitrarily large but must be of the order 1. In this case, the
energy loss of the quark is very small due to the large
non-commutativity.

\vspace{1cm}

\noindent{\bf ACKNOWLEDGMENTS}
\vspace{.5cm}

We would like to thank Shesansu S. Pal for discussion and
collaboration at an early stage of this work. One of us (SR)
would also like to thank Munshi G. Mustafa for discussions. KLP
would like to mention that he misses Alok Kumar and his
inspiration.

\end{document}